\documentclass[10pt]{iopart}
\usepackage{epsfig}
\usepackage{cite}

\begin{document}

\title{Superconductivity in a misfit layered compound (SnSe)$_{1.16}$(NbSe$_2$)}

\author{Hua Bai$^1$, Xiaohui Yang$^1$, Yi Liu$^1$, Meng Zhang$^1$, Mengmeng Wang$^1$, Yupeng Li$^1$, Jiang Ma$^1$, Qian Tao$^1$, Yanwu Xie$^1$, Guanghan Cao$^{1,}$$^{3,}$$^{4}$  Zhu-An Xu$^{1,}$$^{2,}$$^{3,}$$^{4}$\footnote[3]{Corresponding
author. }}

\address{$^1$Department of Physics, Zhejiang University, Hangzhou 310027, P. R. China}

\address{$^2$State Key Laboratory of Silicon Materials, Zhejiang University, Hangzhou 310027, P. R. China}

\address{$^3$Zhejiang California International NanoSystems Institute, Zhejiang University, Hangzhou 310058, P. R. China}

\address{$^4$Collaborative Innovation Centre of Advanced Microstructures, Nanjing University, Nanjing 210093, P. R. China}

\ead{zhuan@zju.edu.cn}
\date{\today}

\begin{abstract}
The large size single crystals of (SnSe)$_{1.16}$(NbSe$_2$) misfit
layered compound were grown and superconductivity with $T_c$ of
3.4 K was first discovered in this system. Powder X-ray
diffraction (XRD) and high resolution transmission electron
microscopy (HRTEM) clearly display the misfit feature between SnSe
and NbSe$_2$ subsystems. The Sommerfeld coefficient $\gamma$
inferred from specific-heat measurements is 16.73 mJ mol$^{-1}$
K$^{-2}$, slightly larger than the usual misfit compounds. The
normalized specific heat jump $\Delta$$C_e$/$\gamma$$T_{\rm c}$ is
about 0.98, and the electron-phonon coupling constant
$\lambda$$_{e-ph}$ is estimated to be 0.80. The estimated value of
the in-plane upper critical magnetic field, $H_{c2}^{ab}$(0), is
about 7.82 T, exceeding the Pauli paramagnetic limit slightly.
Both the specific-heat and $H_{c2}$ data suggest that
(SnSe)$_{1.16}$(NbSe$_2$) is a multi-band superconductor.

\end{abstract}

\maketitle

\setlength{\parskip}{0\baselineskip}

\ioptwocol

\section{INTRODUCTION}
The misfit layered compounds (MLCs) has been more attractive due to their unique layered structure recently.
The formula of MLCs is generally described as $(MX)$$_{1+y}$($TX$$_2$)$_n$ ($y$ = 0.08-0.28, $n$ = 1, 2, 3),
where $M$ = Sn, Pb, Sb, Bi or a lanthanide; $X$ = S, Se or Te; $T$ = Ti, V, Cr, Nb or Ta.
Rocksalt structure monochalcogenides layer $MX$ and several layers of transition metal
dichalcogenides (TMDCs) $TX$$_2$ are connected by van der Waals forces, forming natural heterojunctions.
The $MX$ and $TX$$_2$ structural components have different  rotation symmetries - they have the same
periodicity along one in-plane direction but different periodicity along the other
in-plane direction. Therefore, when they are stacked together, incommensurate factor 1+$y$  appears\cite{Wiegers1996Misfit}.
In the MLCs family, the members of Nb, Ta and Ti dichalcogenides with rocksalt Sn, Pb, Bi and La monochalcogenide layers often exhibit superconductivity\cite{Reefman1990Superconductivity,Hamersm1992Anomalies,Oosawa1992Three,Roesky1993ChemInform,
Rouxel1995Chalcogenide,Nader1997Superconductivity,Nader1997Superconductivity2,Nader1998Structural,Gressier2006Misfit,Giang2010Superconductivity,Luo2016Superconductivity,
Song2016Superconducting,Sankar2018Superconductivity}, with a $T_{\rm c}$ ranging from 0.4 to 5.3 K. Most MLC superconductors are weak-coupling
BCS-type superconductors, and usually have low values of upper critical field ($H_{c2}$) in both inter-plane and in-plane directions\cite{Reefman1991Superconductivity,Nader1996Critical,Nader1997Superconductivity2,Nader1998Critical,Song2016Superconducting,Sankar2018Superconductivity}.

Similar to MLCs, the ferecrystals are another type layered compounds consisting of same $MX$ and $TX$$_2$ layers.
The only difference is that the ferecrystals have extensive random  rotationally disorder between layers but MLCs do
not\cite{Atkins2013Synthesis,Beekman2014Ferecrystals,Westover2014Synthesis,Alemayehu2015Structural,Grosse2017Superconducting}.
Therefore the ferecrystals are not in the long-range crystalline state.
Most MLCs were prepared through traditional high-temperature synthetic techniques,
while most ferecrystals were grown using physical vapor deposition (PVD) or chemical vapor deposition (CVD) techniques.
Some ferecrystals also show superconductivity, such as the ferecrystals of [(SnSe)$_{1.16}$]$_m$(NbSe$_2$)\cite{Alemayehu2015Structural,Grosse2017Superconducting,0953-2048-31-6-065006}.
The $T_{\rm c}$ value is 1.9 K for $m$ = 1, and it decreases with increasing $m$.

Previous study on the powder sample of MLC (SnSe)$_{1.16}$(NbSe$_2$) did not observe superconductivity as temperature down to 4.2 K \cite{Wiegers1991The} and thus no systematic investigation on its properties has been reported.
Here we report the successful crystal growth of large size single crystals of (SnSe)$_{1.16}$(NbSe$_2$) and discovery of superconductivity with onset transition temperature of 3.4 K. The feature of the misfit structure is revealed by the powder X-ray diffraction (XRD) and high resolution transmission electron microscopy (HRTEM) measurements. The specific-heat and $H_{c2}$ data  can be fit within a two-band superconducting model.
The value of $H_{c2}^{ab}$(0) exceeds the Pauli paramagnetic limit slightly which may result from the multi-band nature.

\begin{figure*}[htp]
\centering
\includegraphics[width=7in]{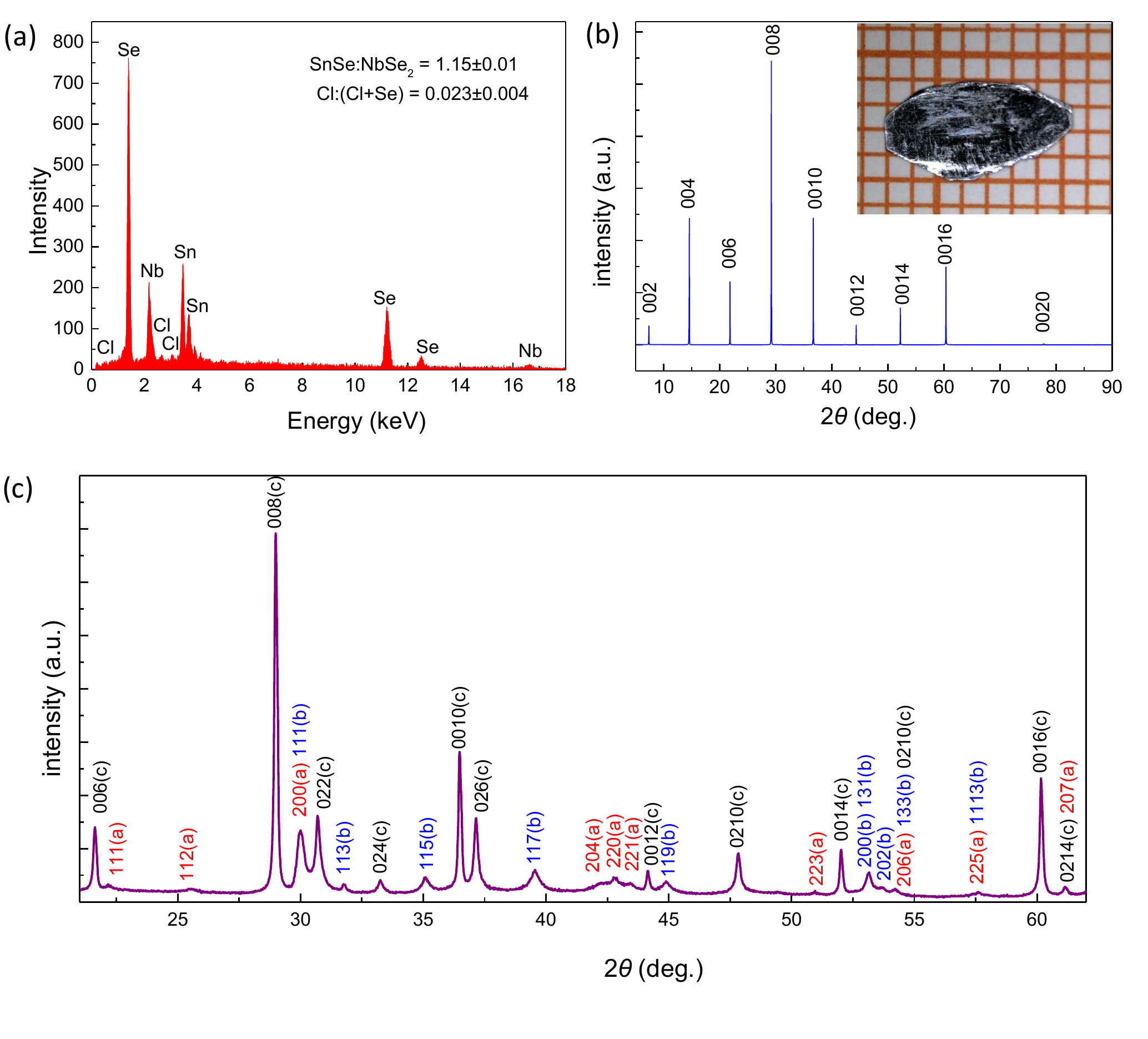}
\caption{\label{Fig1}(a) EDX pattern of (SnSe)$_{1.16}$(NbSe$_2$).
(b) XRD pattern of a (SnSe)$_{1.16}$(NbSe$_2$) single crystal. Inset:
Photograph of the as-grown (SnSe)$_{1.16}$(NbSe$_2$) single crystal.
(c) XRD pattern of (SnSe)$_{1.16}$(NbSe$_2$) powder. ``a'' represents SnSe subsystem,
``b'' represents NbSe$_2$ subsystem, and ``c'' represents common
reflections with indices of the NbSe$_2$ subsystem. }
\end{figure*}

\section{EXPERIMENT DETAILS}

The single crystals of (SnSe)$_{1.16}$(NbSe$_2$) were made via
vapor transport technique using chlorine as the transport agent.
First, Sn(99.999\%), Nb(99.99\%) and Se(99.999\%) powders with the
total mass of 1.5 g were mixed with the ratio of 1.16:1:3.16 and
ground adequately, sealed into an evacuated quartz ampoule, heated
to 900 $^{\circ}$C and kept for 3 days. Subsequently, the mixture
was reground with 0.15 g powders of SnCl$_2$(99.999\%), and sealed
into an evacuated quartz ampoule with a length of 16 cm. The
ampoule was heated for 7 days in a two-zone furnace, where the
temperature of source zone and growth zone were fixed at 900
$^{\circ}$C and 850 $^{\circ}$C respectively. Finally, large
plate-like single crystals of (SnSe)$_{1.16}$(NbSe$_2$) with a
size of several millimeters in the plane direction were obtained,
and their thickness is about 0.1-0.2 mm.

The single crystal XRD and powder XRD data were collected by
a PANalytical X-ray diffractometer (Empyrean) with a Cu K$_\alpha$
radiation and a graphite monochromator at room temperature. The
chemical compositions were determined by energy-dispersive X-ray
spectroscopy (EDX) with a GENESIS4000 EDAX spectrometer.
The HRTEM images were taken at room temperature with an aberration corrected
FEI-Titan G2 80-200 ChemiSTEM. The DC magnetization was measured on a
magnetic property measurement system (MPMS-XL5, Quantum Design),
The specific-heat capacity was measured on a physical properties
measurement system (PPMS-9, Quantum Design), using a relaxation technique.
The electrical properties were measured on an Oxford Instruments-15T
cryostat with a He-3 probe option.

\begin{figure*}[htp]
\begin{center}
\includegraphics[width=7in]{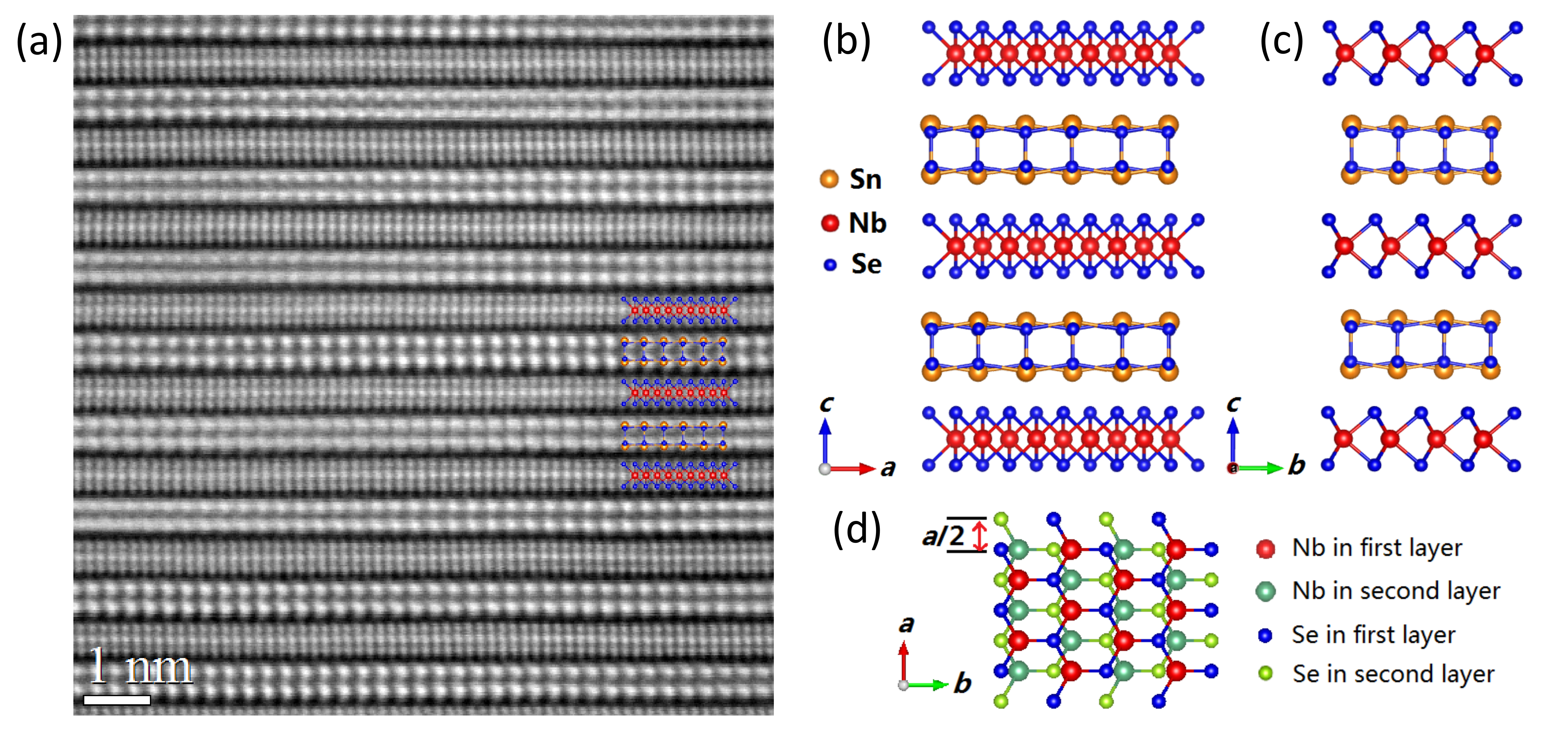}
\end{center}
\caption{\label{Fig2} (a) HRTEM image of a (SnSe)$_{1.16}$(NbSe$_2$)
single crystal from zone axes [010]. Orange, red, and blue colors denote Sn, Nb, and Se atoms.
(b) and (c) The structure of (SnSe)$_{1.16}$(NbSe$_2$).
 (b) Along the b-axis. (c) Along the a-axis.
(d) The illustration of the stack mode of NbSe$_2$ subsystem along the c-axis.
In order to distinguish two layers, the color of Nb atoms in the second layer is changed to green,
and the color of Se atoms in the second layer is changed to cyan.}
\end{figure*}

\section{RESULTS AND DISCUSSION}

Figure \ref{Fig1} (a) shows the EDX pattern of a
(SnSe)$_{1.16}$(NbSe$_2$) single crystal. According to the EDX
result, the ratio of SnSe and NbSe$_2$ is about 1.15(1):1, very
close to the nominal ratio. In addition, a tiny part of Se (about
2.3\%) are substituted by Cl, which is unavoidable due to the
using of chlorine transport agent. We also tried to use iodine as
a transport agent but failed to obtain required crystals. Figure
\ref{Fig1} (b) shows the room temperature XRD pattern of the
single crystal, where all the reflections are (00$l$) peaks. The
inset of Figure \ref{Fig1} (b) shows a typical as-grown single
crystal. The soft and plate-like single crystals are difficult to
grind to powder. We managed to mince them to powder adequately by
blade and obtained the powder XRD data finally, which are showed
in Figure \ref{Fig1} (c). The XRD patterns are almost the same as
reported in the literature\cite{Wiegers1991The}, which can be well
indexed by the SnSe and NbSe$_2$ subsystems. The detailed indices
of the peaks are labelled in Fig.1(c). We calculated the lattice
parameters of two subsystems subsystems respectively by using the
UnitCell program\cite{Holland1997Unit}. The lattice parameters of
two subsystems are shown in Table 1.

\begin{center}
\textbf{Table 1}~~The lattice parameters of (SnSe)$_{1.16}$(NbSe$_2$).\\
\setlength{\tabcolsep}{2.3mm}{
\begin{tabular}{ccccc}
\hline
subsystem& $a$ (\AA)& $b$ (\AA)& $c$ (\AA)& space group\\
\hline
SnSe& 5.940& 5.968& 12.340& Cm2a\\
NbSe$_2$& 3.444& 5.968& 24.680& Fm2m\\
\hline
\end{tabular}}
\end{center}
The structural parameters are very close to
the reported values in the literature\cite{Wiegers1991The}. The
incommensurate factor can be calculated by the formula: 1+$y$ =
2$a_2$/$a_1$ $\approx$ 1.16, where $a_2$ and $a_1$ are lattice
parameters of a-axis for NbSe$_2$ and SnSe subsystems
respectively.

\begin{figure*}[htp]
\begin{center}
\includegraphics[width=7in]{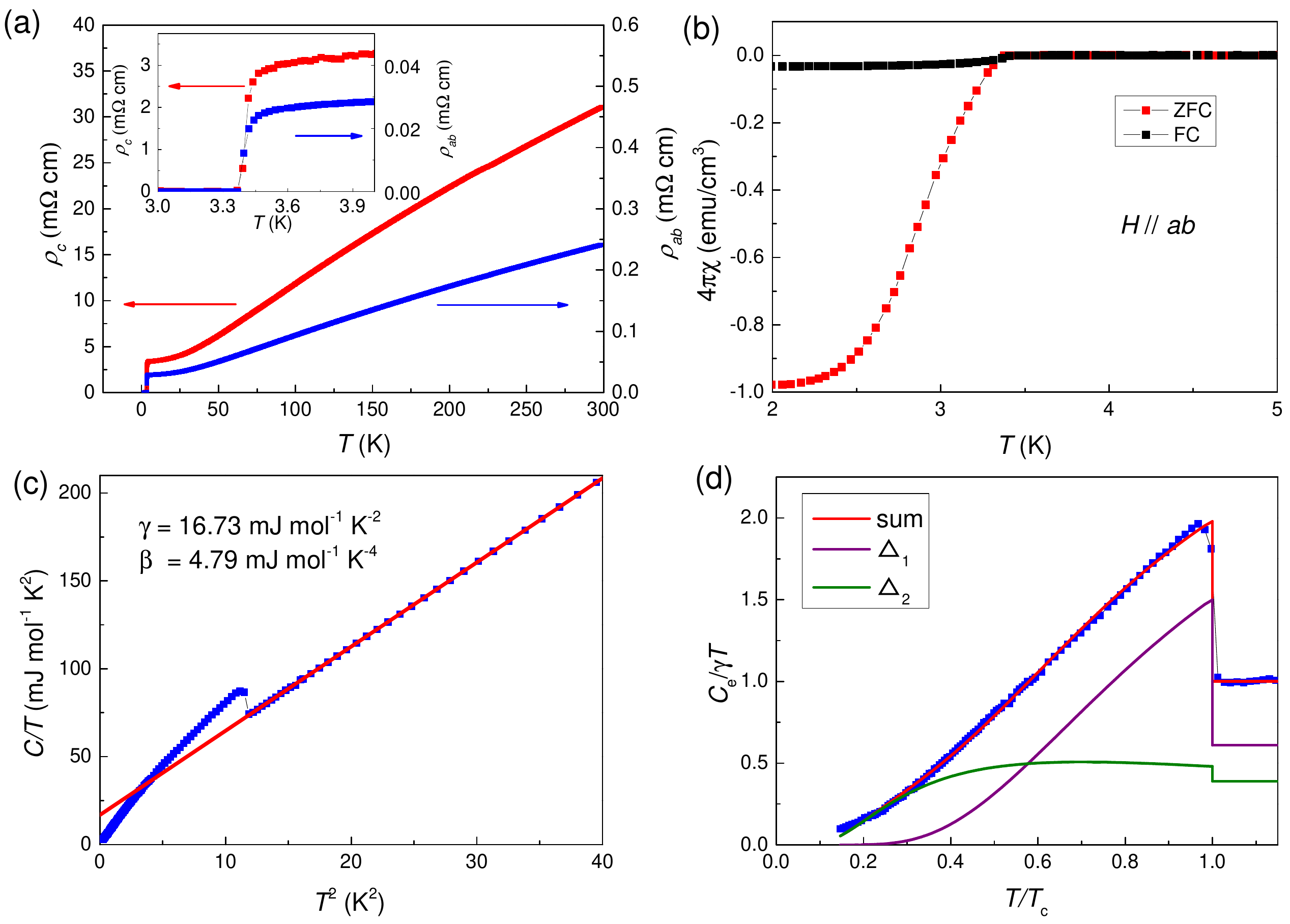}
\end{center}
\caption{\label{Fig3}(a) Temperature dependent out-of-plane and in-plane
resistivity of (SnSe)$_{1.16}$(NbSe$_2$) from 1.5 to 300 K.
Inset: Expanded view of superconducting transition around 3.4 K for both directions.
(b) Temperature dependence of dc magnetic susceptibility ($H//ab$, $H$ = 10 Oe) around $T_{\rm c}$.
(c) Specific heat divided by temperature, $(C/T)$, as a function of $T^2$, and
the red line is a linear fit. (d) $C_e/(\gamma T)$ around $T_{\rm c}$, as a fuction of $T/T_{\rm c}$.
the red line is a fit of two-gap BCS model, the purple line and green line are respective parts of two bands.
}
\end{figure*}

Figure \ref{Fig2} (a) shows the HRTEM image of a
(SnSe)$_{1.16}$(NbSe$_2$) single crystal from zone axes [010]. The
HRTEM image clearly displays regular stacking of SnSe and NbSe$_2$
layers. Furthermore, the different periodicity of two subsystems
along the $b$-axis is revealed. The structure model of
(SnSe)$_{1.16}$(NbSe$_2$) is displayed in Figures \ref{Fig2} (b)
and (c) in two directions. The SnSe layer is a rocksalt like
bi-layer structure and one unit cell contains one bi-layer in the
SnSe subsystem. While the NbSe$_2$ layer is a trigonal sandwich
structure, the same as the layer in 2$Ha$-NbSe$_2$. The stack mode
for the NbSe$_2$ layers along the $c$-axis is shown in figure
\ref{Fig2} (d), one unit cell contains two layers, the green balls
represent Nb atoms and the cyan balls represent Se atoms in the
second layer, and the first layer has a offset of $a$/2 along the
$a$-axis to the second layer, constituting an orthorhombic
pattern. This special stack mode for TMDCs only exists in MLCs.

\begin{figure*}[htp]
\begin{center}
\includegraphics[width=7in]{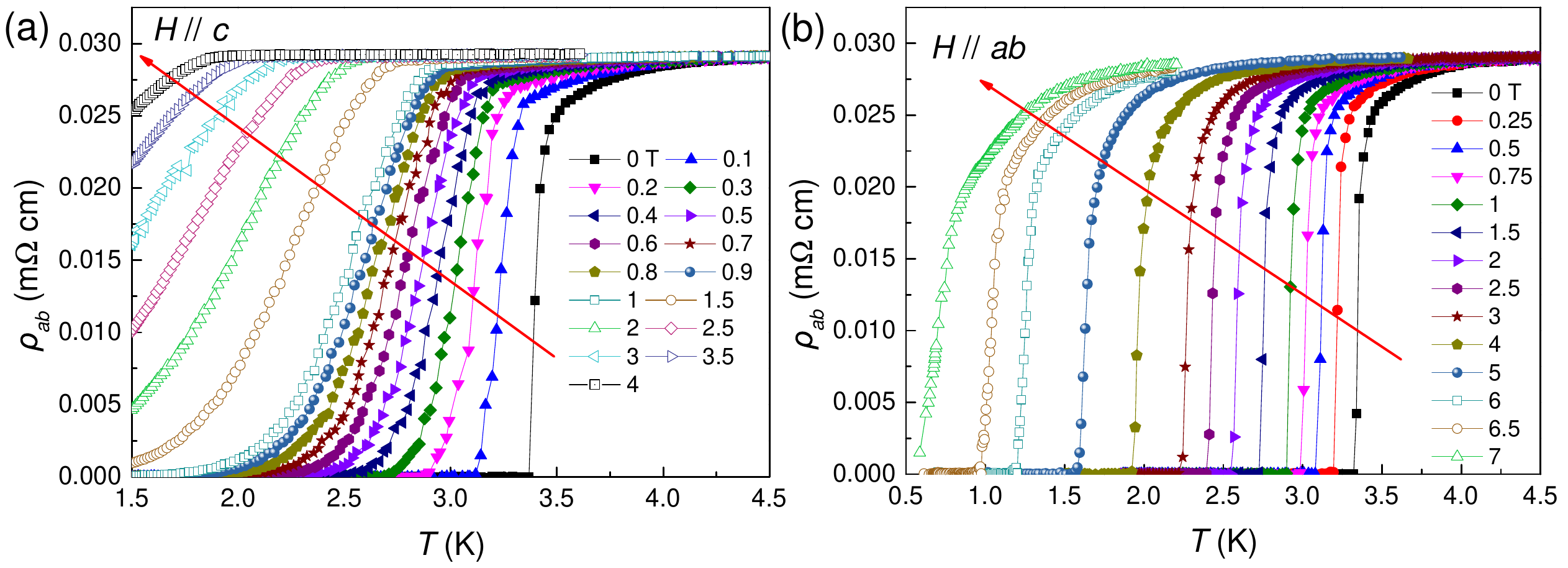}
\end{center}
\caption{\label{Fig4} The in-plane resistivity under different
magnetic fields of (SnSe)$_{1.16}$(NbSe$_2$). (a) $H//c$, up to 4 T. (b) $H//ab$, up to 7 T.}
\end{figure*}

Figure \ref{Fig3} (a) displays the temperature dependence of
out-of-plane and in-plane electrical resistivity for a
(SnSe)$_{1.16}$(NbSe$_2$) single crystal. In both directions,
normal state of (SnSe)$_{1.16}$(NbSe$_2$) shows a metallic
behavior. The inset shows the details of superconducting
transition. The onset superconducting transition temperature
$T_{\rm c}$ (90\% of normal state resistivity) is about 3.4 K for
both directions, with a very narrow transition width of about 0.02
K. The anisotropy of resistivity $\rho_c$/$\rho_{ab}$ is about 128
at 300 K, and exceeds 100 over the whole temperature range. Such a
large anisotropy of resistivity should be related to the highly
anisotropic structure. Figure \ref{Fig3} (b) shows the temperature
dependence of dc magnetic susceptibility under magnetic field of
10 Oe applied in the $ab$-plane, where both zero-field cooling
(ZFC) and field cooling (FC) were measured. The $T_{\rm c}$
determined from magnetic susceptibility is 3.38 K, almost the same
as $T_{\rm c}^{\rm zero}$ in resistivity. The estimated
superconducting shielding volume fraction is very close to 100\%
at 2 K, indicating completely bulk superconductivity in
(SnSe)$_{1.16}$(NbSe$_2$). The demagnetization factor is not
considered due to its puny influence in the $ab$-plane  direction
for a plate-like sample.

Figure \ref{Fig3} (c) shows the low temperature specific-heat data
plotted as $C/T$ versus $T^2$. The specific-heat of normal state above $T_{\rm c}$ contains
two parts, which can be described by the equation $C/T$ = $\gamma$+${\beta}T^2$, where
 $\gamma$ is the Sommerfeld coefficient for electronic specific-heat and
 $\beta$ is the lattice specific-heat coefficient.  From this linear fit,
 we can obtained that $\gamma$ = 16.73 mJ mol$^{-1}$ K$^{-2}$ and $\beta$ = 4.79 mJ mol$^{-1}$ K$^{-4}$.
 Furthermore, the Debye temperature $\Theta$$_D$ can be estimated by the formula
 $\Theta$$_D$ = (12$\pi$$^4nR/5\beta$)$^{1/3}$ = 129.3 K, where $R$ is the gas constant
 and $n$ is the number of atoms per formula unit ($n$ = 5.32).
 In addition, the electron-phonon coupling constant $\lambda$$_{e-ph}$
 can be estimated by the McMillan equation\cite{Mcmillan1968Transition}:
 \begin{equation} %\begin{split}
\label{1}
%\begin{align}
\lambda_{e-ph}=\frac{\mu^*{\rm ln}(\frac{\Theta_D}{1.45Tc})+1.04}{{\rm ln}(\frac{\Theta_D}{1.45Tc})(1-0.62\mu^*)-1.04} %\label{1}
%\end{align}
%\end{split}
\end{equation}
where $\mu^*$ is Coulomb pseudopotential which is often set to 0.15 as the empirical value.
The estimated $\lambda$$_{e-ph}$ = 0.80, suggesting that (SnSe)$_{1.16}$(NbSe$_2$) is in an intermediate coupling range.
The $\lambda$$_{e-ph}$ value is also similar to other MLC superconductors\cite{Sankar2018Superconductivity,Luo2016Superconductivity,Song2016Superconducting}.
An obvious specific-heat jump appears around  $T_{\rm c}$,
which is also a characteristic for superconducting transition.
Figure \ref{Fig3} (d) shows the normalized electronic specific-heat
$C_e$/$\gamma$$T$, as a function of reduced temperature $t$ = $T/T_{\rm c}$.
The $C_e$ data cannot be fitted with the one-gap BCS model: $C_e$ = $C_0$exp(-$\Delta$/$k_B$$T$).
 Therefore, we adopted a two-gap BCS model which is often applied to multi-band superconductors\cite{Wang2013Multiband,Bhoi2016Interplay,Bouquet2001Phenomenological,Yang2012Comparative,Huang2007Experimental}
. In this model, $C_e$ is the sum of contributions
from two bands where each band follows one-gap BCS model. By this fit, we obtained that 2$\Delta$$_1$ = 4.52 $k_B$$T_{\rm c}$
while 2$\Delta$$_2$ = 1.4 $k_B$$T_{\rm c}$, and the weight $\gamma$$_1$/$\gamma$$_2$ = 60\%:40\%.
Therefore, one gap exceed the BCS value of 3.53 $k_B$$T_{\rm c}$, while the other gap is smaller than it.
2$H$-NbSe$_2$ is also a multi-band superconductor. In 2$H$-NbSe$_2$,
2$\Delta$$_1$ = 4.5 $k_B$$T_{\rm c}$, 2$\Delta$$_2$ = 2.6 $k_B$$T_{\rm c}$,
and the weight $\gamma$$_1$/$\gamma$$_2$ = 80\%:20\%\cite{Huang2007Experimental}.
We can regard (SnSe)$_{1.16}$(NbSe$_2$) as an intercalated NbSe$_2$.
Therefore,  it is reasonable that (SnSe)$_{1.16}$(NbSe$_2$) and 2$H$-NbSe$_2$ have the similar $\Delta$$_1$,
but the former has a smaller $\Delta$$_2$. The weights for two gaps are also different.
MLCs always have charge transfer from $MX$ layers to $TX_2$ layers\cite{Wiegers1996Misfit,Rouxel1995Chalcogenide,Giang2010Superconductivity,Yao2018Charge}.
We speculate that the charge transfer changes the Fermi surface of NbSe$_2$
in (SnSe)$_{1.16}$(NbSe$_2$), leading to the differences of gaps between the two compounds.
Moreover, the normalized specific heat jump $\Delta$$C_e$/$\gamma$$T_{\rm c}$ is estimated
to be 0.98, close to the value in (SnS)$_{1.15}$(TaS$_2$)(0.812)\cite{Sankar2018Superconductivity} and
(SnSe)$_{1.18}$(TiSe$_2$)$_2$ (0.88)\cite{Song2016Superconducting}, but lower than
(Pb$_{0.8}$Sn$_{0.2}$Se)$_{1.16}$(TiSe$_2$)$_2$ (1.38)\cite{Luo2016Superconductivity}.

Figures \ref{Fig4} (a) and (b) show the temperature dependence of
in-plane electrical resistivity around $T_{\rm c}$ under different magnetic fields of a
(SnSe)$_{1.16}$(NbSe$_2$) single crystal with field applied parallel to the
$c$-axis and in the $ab$-plane respectively. For $H//c$, the superconducting
transition width increases significantly with increasing fields, while for $H//ab$,
the resistivity curves shift parallel down towards the low temperature with increasing fields.

\begin{figure}[htp]
\begin{center}
\includegraphics[width=3.3in]{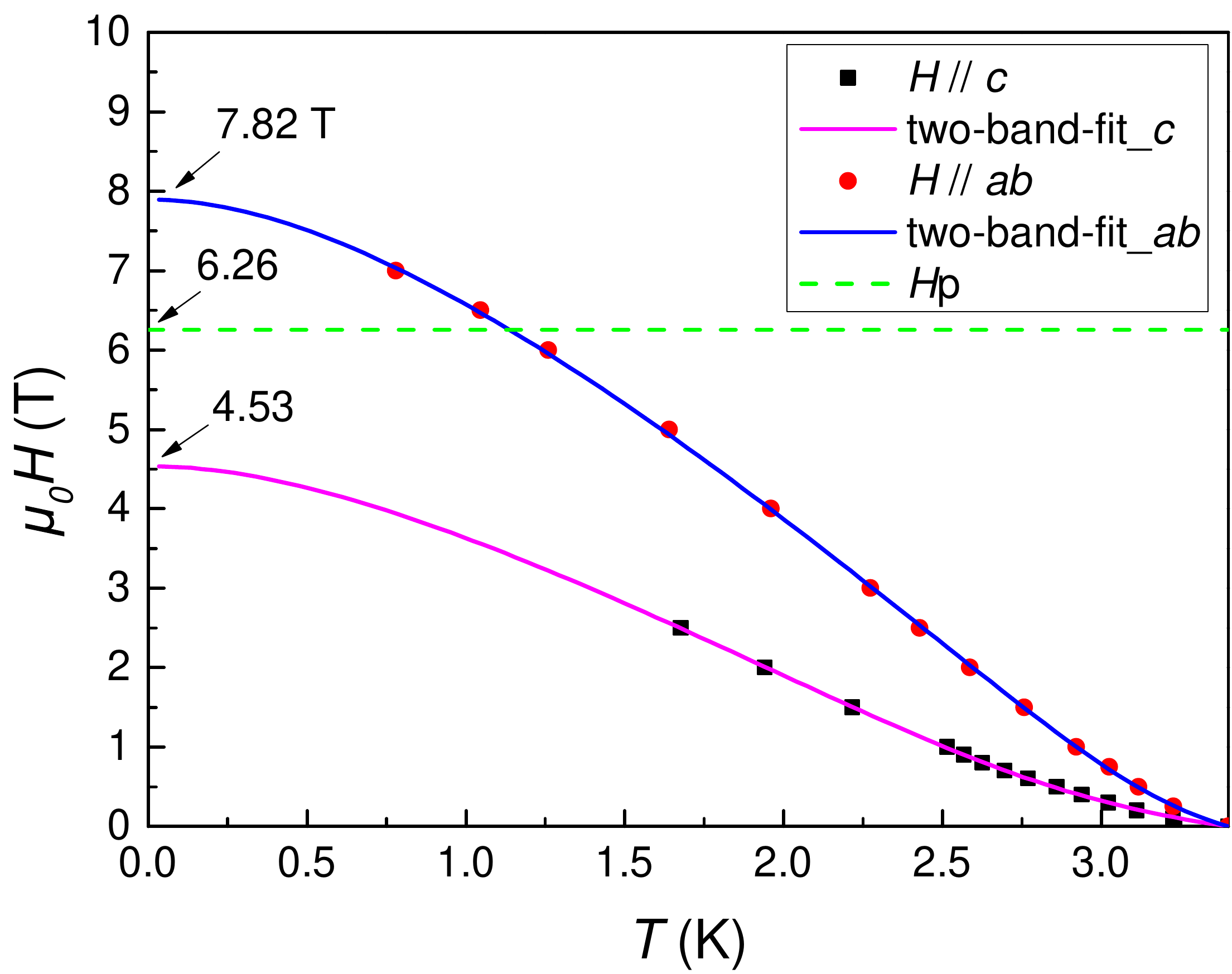}
\end{center}
\caption{\label{Fig5} The upper critical field determined by the $T_{\rm c}^{\rm mid}$ and the two-band fits for both field directions. The dash
 cyan line dictates the Pauli paramagnetic limit.}
\end{figure}

We summarize the temperature dependent $H_{c2}$ determined by
$T_{\rm c}^{\rm mid}$ (50\% of normal state resistivity) in Figure
\ref{Fig5}. For both directions, the $H_{c2}$ curves exhibit a
positive curvature close to $T_{\rm c}$. The similar phenomenon
were also observed in other multi-band
superconductors\cite{Shulga1998Upper,S2001Magnetoresistivity,Hunte2008Two,Wang2013Multiband,Bhoi2016Interplay}.
Therefore we adopt a two-band model for
$H_{c2}$\cite{Gurevich2003Enhancement}:
%\begin{equation} %\begin{split}

%\begin{alignd}
\begin{eqnarray}
a_0[{\rm ln}t+U(h/t)][{\rm ln}t+U(\eta h/t)] + \nonumber\\
a_1[{\rm ln}t+U(h/t)] + a_2[{\rm ln}t+U(\eta h/t)] = 0
 %\label{1}
\end{eqnarray}
%\begin{alignd}
%\end{equation}
where $t=T/T_{\rm c}$ is the reduced temperature.
$a_0$ = 2($\lambda_{11}$$\lambda_{22}$-$\lambda_{12}$$\lambda_{21}$),
$a_1$ = 1+($\lambda_{11}$-$\lambda_{22}$)/$\lambda_{0}$,
$a_2$ = 1-($\lambda_{11}$-$\lambda_{22}$)/$\lambda_{0}$,
$\lambda_{0}$ = [($\lambda_{11}$-$\lambda_{22}$)$^2$+4$\lambda_{12}$$\lambda_{21}$]$^{\frac{1}{2}}$,
$h$ = $H_{c2}$$D_1$/2$\phi_0$$T$, $\eta$ = $D_1$/$D_2$,
$U(x)$ = $\Psi$$(x+1/2)$-$\Psi$$(x)$, the $\Psi$$(x)$ is digamma function.
$\lambda_{11}$ and $\lambda_{22}$ are the intraband BCS
coupling constants, where $\lambda_{12}$ and $\lambda_{21}$ are the interband BCS coupling constants,
and $D_1$ and $D_2$ are the in-plane diffusivity of each band. The fit parameters are listed in Table 2.
The fitting parameters reveal the differences between two bands.
For the first band, the intraband coupling constant is greater than 1,
which is in a strong-coupling range. Furthermore,
the intraband coupling is much stronger than the interband coupling.
While in the second band, the intraband coupling constant is still in
a weak-coupling range, and is similar to interband coupling.
This fitting results are consistent with the two-gap fit in specific-heat,
proving the multi-band superconductivity in (SnSe)$_{1.16}$(NbSe$_2$) qualitatively.
In addition, this fit yields $\mu$$_0H_{c2}^c$(0) = 4.53 T and
$\mu$$_0H_{c2}^{ab}$(0) = 7.82 T. The Pauli paramagnetic limit ($H_P$)
for the upper critical field is $\mu$$_0H_P$ = 1.84 $T_{\rm c}$ =
6.26 T. Therefore the value of $\mu$$_0H_{c2}^{ab}$(0) exceed $H_P$
slightly. As mentioned above, most MLC superconductors have low values of
$H_{c2}$ in both directions which are below $H_P$, hence the values of $H_{c2}$ in
(SnSe)$_{1.16}$(NbSe$_2$) are unusually large. We suggest that the multi-band effects may enhance $H_{c2}$\cite{Hunte2008Two,Bhoi2016Interplay}.
The anisotropy factor $\gamma$ = $H_{c2}^{ab}(0)$/$H_{c2}^c(0)$ $\approx$ 1.73,
which is within the range of others MLC superconductors (1.4 - 5.29)
\cite{Reefman1991Superconductivity,Nader1997Superconductivity2,Nader1998Critical,Song2016Superconducting,Sankar2018Superconductivity}.

\begin{center}
\textbf{Table 2}~~The fit parameters of two-band model.\\
\setlength{\tabcolsep}{2.5mm}{
\begin{tabular}{ccccccc}
\hline
direction& $\lambda_{11}$& $\lambda_{22}$& $\lambda_{12}$& $\lambda_{21}$& $\eta$& $D_1$\\
\hline
$c$& 1.09& 0.44& 0.17& 0.65& 20.4& 0.31\\
$ab$& 1.18& 0.60& 0.08& 0.62& 20.4& 0.27\\
\hline
\end{tabular}}
\end{center}

\section{CONCLUSION}
In summary, we have successfully grown large single crystals of
MLC (SnSe)$_{1.16}$(NbSe$_2$) and discovered superconductivity
with  $T_{\rm c}$ of 3.4 K. The details of the misfit structure
were revealed by the powder XRD and HRTEM measurements.
Superconducting shielding volume fraction is close to 100\%,
confirming the bulk superconductivity. The Sommerfeld coefficient
$\gamma$ is 16.73 mJ mol$^{-1}$ K$^{-2}$, and
 $H_{c2}$(0) is estimated to be 4.53 T and 7.82 T for the out-plane and in-plane directions respectively. Especially for
the in-plane direction, the value of $H_{c2}$(0) exceeding the
Pauli limit $H_P$ slightly. The electron-phonon coupling constant
$\lambda$$_{e-ph}$ = 0.80. Both the specific-heat and $H_{c2}$
data suggest that (SnSe)$_{1.16}$(NbSe$_2$) should be a multi-band
BCS superconductor.

\section{Acknowledgments}
 This work was supported by the National Basic Research Program of China
(Grant Nos. 2014CB92103) and the National Key R\&D Projects of
China (Grant No. 2016FYA0300402 and 2016FYA0300204), the National Science Foundation
of China (Grant No. 11774305), and the Fundamental
Research Funds for the Central Universities of China.

\section{References}
\providecommand{\newblock}{}

\end{document}